\documentclass[pra,twocolumn,showpacs,amsfonts,amsmath,floatfix]{revtex4}
\usepackage{graphicx}
\newcommand{\Journal}[4]{#1 {\bf #2}, #3 (#4)}
\newcommand{\PRL}{Phys. Rev. Lett.}

\newcommand{\Science}{Science}
\newcommand{\PLA}{Phys. Lett. A}
\begin{document}
\title{Two super Tonks-Girardeau states of a trapped 1D spinor Fermi gas}
\author{M. D. Girardeau}
\email{girardeau@optics.arizona.edu}
\affiliation{College of Optical Sciences, University of Arizona, Tucson, AZ 85721, USA}
\date{\today}
\begin{abstract}
A harmonically trapped ultracold 1D spinor Fermi gas with a strongly attractive 1D even-wave interaction induced by a 3D Feshbach resonance is studied. It is shown that it has two different super Tonks-Girardeau (sTG) energy eigenstates
which are metastable against collapse in spite of the strong attraction, due to their close connection with 
1D hard sphere Bose gases which are highly excited gas-like states. 
One of these sTG states is a hybrid between an sTG gas with strong $(\uparrow\downarrow)$ attractions and an ideal Fermi gas with no
$(\uparrow\uparrow)$ or $(\downarrow\downarrow)$ interactions, the sTG component being an exact analog of the recently observed
sTG state of a 1D ultracold Bose gas. It should be possible to create it experimentally by a sudden switch of the $(\uparrow\downarrow)$
interaction from strongly repulsive to strongly attractive, as in the recent Innsbruck experiment on the bosonic sTG gas.
The other is a trapped analog of a recently predicted sTG state which is an ultracold gas of strongly bound 
$(\uparrow\downarrow)$ fermion dimers which behave as bosons with a strongly attractive boson-boson interaction leading to sTG behavior.
It is proved that the probability of a transition from the ground state for strongly repulsive interaction
to this dimer state under a sudden switch from strongly repulsive to strongly attractive interaction is $\ll 1$, contrary to a previous suggestion.
\end{abstract}
\maketitle
If an ultracold atomic vapor is confined in a de Broglie wave guide with
transverse trapping so tight and temperature so low that the transverse
vibrational excitation quantum is larger than available
longitudinal zero point and thermal energies, the effective dynamics becomes
one-dimensional (1D) \cite{Ols98,PetShlWal00}. 3D Feshbach resonances \cite{Rob01}
allow tuning to the neighborhood of
1D confinement-induced resonances \cite{Ols98,BerMooOls03} where the 1D interaction is very
strong, leading to strong short-range correlations, breakdown of
effective-field theories, and emergence of highly-correlated $N$-body
ground states. In the bosonic case with zero-range repulsion $g_B\delta(x_j-x_\ell)$
with coupling constant $g_B\to +\infty$, the
Tonks-Girardeau (TG) gas, the exact
$N$-body ground state was determined in 1960 by a
Fermi-Bose (FB) mapping to an ideal Fermi gas \cite{Gir60}, leading to
``fermionization'' of many properties of this Bose system, as recently
confirmed experimentally \cite{Par04,Kin04}. It is now known
\cite{CheShi98,GirOls03,GirNguOls04} that the FB mapping
is of much greater generality; when supplemented by an inversion and sign
change of the coupling constant, it provides a mapping between
the $N$-body energy eigenstates of a 1D Bose gas with delta-function
interactions $g_{B}\delta(x_j-x_\ell)$ of any strength [Lieb-Liniger (LL) gas \cite{LieLin63}] and those
of a spin-aligned Fermi gas.

A generalization of the FB mapping 
to a 1D spinor Fermi gas is relevant to a recent prediction of a super Tonks-Girardeau (sTG)
state in such a system \cite{CheGuaYinGuaBat09}, and will be employed herein. The model consists of an ultracold gas of
fermionic atoms in an effectively 1D trap, in two different hyperfine states which are conveniently labelled as $\uparrow$ (spin up) and
$\downarrow$ (spin down). The previous analysis \cite{CheGuaYinGuaBat09} assumed a ring trap (periodic boundary conditions), but all experiments on ultracold  gases with tight transverse trapping (1D regime) use a linear geometry with a longitudinal harmonic trap
potential. The Hamiltonian generalizes that of \cite{CheGuaYinGuaBat09} by adding a spin-independent harmonic trapping potential:
\begin{equation}\label{H}
\hat{H}_{\text{F}}=\sum_{j=1}^N\left(-\frac{\hbar^2}{2m}\frac{\partial^2}{\partial x^2}+\frac{m\omega^2}{2}x_j^2\right)
+g_{\text{F}}\hspace{-0.3cm}\sum_{1\le j<\ell\le N}\delta(x_j-x_\ell)\ .
\end{equation}
Since spin-up atoms are distinguishable from spin-down atoms, both 3D s-wave and p-wave scattering are allowed. The case relevant to
generation of a sTG state is that of strong s-wave scattering due to a 3D s-wave Feshbach resonance, which leads in 1D to a 
confinement-induced even-wave resonance \cite{Ols98,BerMooOls03} and a LL delta function interaction \cite{LieLin63}. It acts
between atoms $j$ and $\ell$ only if the wave function is symmetric under spatial exchange $(x_j\leftrightarrow x_\ell)$, in which case
fermionic antisymmetry under combined space-spin exchange $(x_j,\sigma_j)\leftrightarrow(x_\ell,\sigma_\ell)$ demands that the spins
be antiparallel, $\sigma_j=\uparrow$ and $\sigma_\ell=\downarrow$ or vice versa. On the other hand, the delta function interaction is
cancelled in the case that the spin orientations are $\uparrow\uparrow$ or $\downarrow\downarrow$, in which case the required spatial
antisymmetry requires that the wave function vanish when $x_j=x_\ell$. 

If $g_{\text{F}}$ is large and negative, two quite different sTG states can occur.
One is a hybrid between an sTG gas with strong $(\uparrow\downarrow)$ attractions and an ideal Fermi gas with no
$(\uparrow\uparrow)$ or $(\downarrow\downarrow)$ interactions, the sTG component being an exact analog of the sTG state
in an ultracold 1D Bose gas predicted in \cite{AstBluGioGra04,BatBorGuaOel05,AstBorCasGio05} and 
recently created by the Innsbruck group \cite{Haletal09}. 
It should be possible to create it experimentally by a sudden switch of the $(\uparrow\downarrow)$
interaction from strongly repulsive to strongly attractive, as in \cite{Haletal09}.
The other sTG state is a trapped analog of a recently predicted \cite{CheGuaYinGuaBat09} sTG state which is an ultracold gas of strongly bound 
$(\uparrow\downarrow)$ fermion dimers which behave as bosons with a strongly attractive boson-boson interaction leading to sTG behavior.
Both of these sTG states will be studied herein.

Consider first the case $N=2$. The analysis closely parallels that for the spinless Bose gas \cite{GirAst09}. The LL interaction is
$g_{F}\delta(x_1-x_2)$, the harmonic trap potential is $m\omega^{2}(x_1^{2}+x_2^{2})/2$, the wave function for the center of mass (c.m.) coordinate $X=(x_1+x_2)/2$ is  $\psi_{\text{c.m.}}=\exp[-X^2/x_{\text{osc}}^2]$ where $x_{\text{osc}}=\sqrt{\hbar/m\omega}$, 
and its energy is $E_{\text{c.m.}}=\hbar\omega/2$ assuming that the c.m. mode is unexcited. The form of the relative wave function 
$\phi(x)$, with $x=x_{1}-x_{2}$, now depends on whether the spins $\sigma_1$ and $\sigma_2$ are parallel or antiparallel.
In the parallel case the relative wave function $\phi(x)$ is antisymmetric so that the LL interaction is cancelled, and $\phi$
reduces to a trapped harmonic oscillator eigenstate odd in $x$, the product of the ground state Gaussian and a Hermite polynomial
of odd order. In the antiparallel case $\phi(x)$ is symmetric and is identical with the solution for spinless bosons
\cite{Fra03,TemSolSch08,GirAst09}. The solutions are analytic continuations of the Hermite-Gaussians to 
nonintegral quantum number $\nu$, and are parabolic cylinder function $D_{\nu}(q)$ where $x=qx_{\text{osc}}$ and the allowed values of $\nu$ are solutions of a transcendental equation
$\Gamma(\frac{1}{2}-\frac{1}{2}\nu)/\Gamma(-\frac{1}{2}\nu)=-\lambda$
where $\lambda$ is a dimensionless coupling constant $\lambda=g_F/2^{3/2}\hbar\omega x_{\text{osc}}$ \cite{Fra03}. This gives the solution
for $q\ge 0$, while for $q<0$ it is $D_{\nu}(|q|)$, since $\phi(x)$ is even. For $\lambda_F\to+\infty$ the solution reduces to
the first excited harmonic oscillator state $\phi(x)=|q|e^{-q^2/2}$ \cite{GirWriTri01}, the usual TG gas ground state with a cusp at
$q=0$ due to the point hard core interaction. For $\lambda\to -\infty$ this wave function is still an exact energy eigenstate,
but it is highly excited, the ground state being a collapsed state which is an analog, for the trapped system, of
McGuire's cluster state \cite{McG64}. It is an even solution also expressible
in terms of a $D_\nu$, but one whose energy approaches $-\infty$ as $g_F\to -\infty$ ($a_{\text{1D}}\to 0+$); see Fig. 5 of \cite{Fra03}.
For $a_{\text{1D}}\to 0+$ it is well approximated by
$\psi_{B0}\approx\exp(-|x_1-x_2|/a_{\text{1D}})\exp[-(x_1^2+x_2^2)/2x_{\text{osc}}^2]$ where $a_{\text{1D}}\ge 0$ is the 1D 
scattering length. $a_{\text{1D}}$ vanishes as $|\lambda|\to\infty$ and increases with decreasing $|\lambda|$ 
\cite{Ols98,Fra03,TemSolSch08}, and at the same time the node moves from the origin to a position $x_{\text{node}}$
which is close to $a_{\text{1D}}$ when $|\lambda|\gg 1$ but slightly smaller, the explicit expression being given in \cite{GirAst09}. 
As in the case of spinless bosons, the wave function for $|x|>x_{\text{node}}$ is identical, apart from normalization, 
with that of a 1D hard sphere Bose gas \cite{Gir60} with hard sphere diameter $x_{\text{node}}$ \cite{GirAst09}.

Generalizing to $N>2$, denote the $N$-atom wave functions by $\psi_F(x_1,\sigma_1;\cdots;x_N,\sigma_N)$. 
Consider first the case $\lambda\to+\infty$. In this limit the wave functions are required to vanish if $x_j=x_\ell$ provided that  
$\sigma_j=\uparrow\ ,\ \sigma_\ell=\downarrow$ or vice versa, and if $\sigma_j=\sigma_\ell$ they vanish automatically at
$x_j=x_\ell$ by antisymmetry. Furthermore, if the spatial coordinates $(x_1,\cdots,x_N)$ are all different, then the interaction
vanishes independently of $(\sigma_1,\cdots,\sigma_N)$ and the wave functions are kinetic energy eigenstates which are symmetric
under $x_j\leftrightarrow x_\ell$ exchange if $\sigma_j\ne\sigma_\ell$ but antisymmetric if $\sigma_j=\sigma_\ell$. These requirements
are all satisfied by wave functions of the form
\begin{equation}\label{TG-ideal Fermi}
\psi_F=M(x_1,\sigma_1;\cdots;x_N,\sigma_N)\psi_\text{ideal}
\end{equation}
where the spin-dependent Fermi-Fermi mapping function $M$, which maps the strongly interacting spinor Fermi gas to a \emph{spinless ideal} Fermi gas, is
\begin{eqnarray}
&&M(x_1,\sigma_1;\cdots;x_N,\sigma_N)=\prod_{1\le j<\ell\le N}\alpha(x_j,\sigma_j;x_\ell,\sigma_\ell)\nonumber\\
&&\alpha(x_j,\sigma_j;x_\ell,\sigma_\ell)=(\delta_{\sigma_j\uparrow}\delta_{\sigma_\ell\downarrow}
-\delta_{\sigma_j\downarrow}\delta_{\sigma_\ell\uparrow})\text{sgn}(x_j-x_\ell)\nonumber\\
&&\hspace{2.5cm}+\delta_{\sigma_j\uparrow}\delta_{\sigma_\ell\uparrow}+\delta_{\sigma_j\downarrow}\delta_{\sigma_\ell\downarrow}\ ,
\end{eqnarray}
the signum function $\text{sgn}(x)$ is $+1$ ($-1$) if $x>0$ ($x<0$), 
and $\psi_\text{ideal}$ is an energy eigenstate of the trapped 1D ideal gas of ``spinless" fermions, a Slater
determinant of $N$ different harmonic oscillator orbitals. Starting from this ideal Fermi gas, this mapping generates the required
TG-gas cusps in the singlet (spin-antisymmetric, space-symmetric) channel of the strongly interacting spinor Fermi gas, but no
interaction cusps in the triplet (spin-symmetric, space-antisymmetric) channels. Here we are particularly interested in the  
ground state $\psi_{F0}$, which is mapped from the ground state $\psi_{\text{ideal}0}$ of the trapped ideal gas, a 
Slater determinant of the lowest $N$ single-particle eigenfunctions $\phi_n$ of the harmonic oscillator (HO): 
\begin{equation}
\psi_{\text{ideal}0}(x_{1},\cdots,x_{N})=\frac{1}{\sqrt{N!}}
\det_{(n,j)=(0,1)}^{(N-1,N)}\phi_{n}(x_{j})\ .
\end{equation}
The HO orbitals are
\begin{equation}
\phi_{n}(x)= \frac{1}
{\pi^{1/4}x_{osc}^{1/2}\sqrt{2^{n}n!}}e^{-Q^{2}/2}H_{n}(Q)
\end{equation}
with $H_n(Q)$ the Hermite polynomials and
$Q=x/x_{osc}$, where $x_{osc}=\sqrt{\hbar/m\omega}$
is the harmonic oscillator width.
The ground state is a van der Monde determinant \cite{GirWriTri01}
\begin{eqnarray}
\det_{(n,j)=(0,1)}^{(N-1,N)}H_{n}(x_{j})
& = & 2^{N(N-1)/2}\det_{(n,j)=(0,1)}^{(N-1,N)}(x_{j})^{n} \nonumber\\
& = & 2^{N(N-1)/2}\prod_{1\le j<\ell\le N}(x_\ell-x_j)\ ,
\end{eqnarray}
yielding an exact analytical expression
of Bijl-Jastrow pair product form for the $N$-fermion ground state:
\begin{eqnarray}\label{N>2_ground}
&&\psi_{F0}(x_{1},\sigma_1;\cdots;x_{N},\sigma_N)
=C_{N}\left[\prod_{i=1}^{N}e^{-Q_{i}^{2}/2}\right]\nonumber\\
&&\times\prod_{1\le j<\ell\le N}\alpha(x_j,\sigma_j;x_\ell,\sigma_\ell)(x_j-x_\ell)
\end{eqnarray}
with normalization constant
\begin{equation}
C_{N}=2^{N(N-1)/4}\left (\frac{1}{x_{osc}} \right )^{N/2}
\left[N!\prod_{n=0}^{N-1}n!\sqrt{\pi}\right]^{-1/2}\ .
\end{equation}
This TG-ideal Fermi gas hybrid is the \emph{exact ground state} in the TG limit $\lambda=+\infty$.

Suppose that now $\lambda$ is changed instantaneously to $-\infty$. Then the state (\ref{N>2_ground}) is still an exact energy eigenstate, since $\hat{H}$ commutes with the mapping function $M$ except at unlike-spin collision points 
$x_j=x_\ell$, where the wave function vanishes, and the state (\ref{N>2_ground}) maps to the ideal Fermi gas \cite{Gir60,CheShi98}
as in the bosonic case \cite{Gir60,CheShi98,GirAst09}. In fact, this is an exact energy
eigenstate in the limit $|\lambda|\to\infty$ even in the dissipative case where $\lambda$ is complex, since the wave function vanishes
at contact in that limit \cite{Dur09}, validating the FB mapping. This follows from the LL contact condition \cite{LieLin63} 
$2[\partial\psi/\partial x_{jk}]_{x_{jk}=0+}=(mg_F/\hbar^2)\psi(0)$ with a cusp (derivative sign change) at the origin. 
Assuming $\psi$ normalized so that $[\partial\psi/\partial x_{jk}]_{x_{jk}=0+}=-[\partial\psi/\partial x_{jk}]_{x_{jk}=0-}=1$, 
this leads to the 
Taylor expansion $\psi=(2\hbar^2/mg_F)+|x_{jk}|+\cdots$ about $x_j=x_k$, so $\psi$ vanishes at contact when $g_F\to\infty$. 
For $\lambda\ll -1$ such a state is 
highly excited. The much lower ground state is a hybrid of a totally collapsed McGuire cluster state \cite{McG64} for antiparallel spins and an ideal Fermi gas ground state for parallel spins, and Taylor expansion fails in the limit $\lambda\to -\infty$
($a_{\text{1D}}\to 0$), where the derivative of $e^{-|x_j-x_k|/a_{\text{1D}}}$ diverges.
Although the exact ground state is not known for $N>2$ and $-\infty<\lambda\ll -1$, one expects that it will
have the form of Eq. (3) of \cite{GirAst09} for pairs with antiparallel spins, while for pairs with parallel spins it will be an ideal
Fermi gas ground state. One therefore expects that it will be well approximated by
\begin{eqnarray}\label{McGuire-ideal Fermi}
&&\psi_0\approx\prod_{j=1}^N\exp\left(-\frac{x_j^2}{2x_{\text{osc}}^2}\right)\nonumber\\
&&\times\hspace{-.4cm}\prod_{1\le j<\ell\le N}[(\delta_{\sigma_j\uparrow}\delta_{\sigma_\ell\downarrow}
-\delta_{\sigma_j\downarrow}\delta_{\sigma_\ell\uparrow})\exp\left(-\frac{|x_j-x_l|}{a_{\text{1D}}}\right)\nonumber\\
&&+(\delta_{\sigma_j\uparrow}\delta_{\sigma_\ell\uparrow}+\delta_{\sigma_j\downarrow}\delta_{\sigma_\ell\downarrow})(x_j-x_\ell)]
\label{eq:McGuire}
\end{eqnarray}
In fact, the LL contact conditions are satisfied exactly at each collision point $x_j=x_\ell$.

Suppose that $\lambda$ is switched from large positive to large negative values rapidly enough that the sudden approximation is valid. Then the initial wave function will be nearly equal to the state of Eqs. (\ref{TG-ideal Fermi}) and (\ref{N>2_ground}), and the 
wave function after the switch will be a superposition of all eigenstates of the system with the given negative $\lambda$ value,
$\psi_\lambda(t)=\sum_\alpha\langle\psi_{\lambda\alpha}|\psi_{F0}\rangle\psi_{\lambda\alpha}e^{-iE_{\lambda\alpha}t/\hbar}$.
The dominant term in $\psi_\lambda$ will be an sTG state which reduces to (\ref{N>2_ground}) as $\lambda\to -\infty$.
The obvious generalization of the N=2 sTG wave function to arbitrary $N$ differs
from Eqs. (\ref{TG-ideal Fermi}) and (\ref{N>2_ground}) through replacement of the ideal Fermi gas factor $(x_j-x_\ell)$ by the $N=2$ solution $D_\nu$ for antiparallel spins, while retaining the ideal Fermi gas form for parallel spins:
\begin{equation}\label{psiFnu}
\psi_{F\nu}=\left[\prod_{1\le j<\ell\le N}\beta_\nu(x_j,\sigma_j;x_\ell,\sigma_\ell)\right]
\prod_{j=1}^N\exp\left(-\frac{x_j^2}{2x_{\text{osc}}^2}\right)\
\end{equation}
where
\begin{eqnarray}
&&\beta_\nu(x_j,\sigma_j;x_\ell,\sigma_\ell)=
(\delta_{\sigma_j\uparrow}\delta_{\sigma_\ell\uparrow}+\delta_{\sigma_j\downarrow}\delta_{\sigma_\ell\downarrow})(x_j-x_\ell)\nonumber\\
&&\hspace{1.5cm}+(\delta_{\sigma_j\uparrow}\delta_{\sigma_\ell\downarrow}
-\delta_{\sigma_j\downarrow}\delta_{\sigma_\ell\uparrow})D_\nu(|q_{j\ell}|)e^{q_{j\ell}^{2}/4}
\end{eqnarray}
and $q_{j\ell}=(x_j-x_\ell)/x_{\text{osc}}$. It satisfies the contact
conditions exactly. One expects the existence of a highly excited gas-like sTG state with nodes only at a nearest neighbor separation
$|x_{j\ell}|=x_{\text{node}}$ when the spins are antiparallel, as in the Bose case \cite{GirAst09},
where $x_{\text{node}}$ increases with decreasing
$|\lambda|$, is very close to $a_{1D}$ for $|\lambda|\gg 1$, and goes to zero along with $a_{1D}$ in the TG limit $|\lambda|\to\infty$.
For all $N\ge 2$ the approximate wave functions (\ref{psiFnu}) have exactly these properties, vanishing only at
$|x_{j\ell}|=x_{\text{node}}$ when the spins are antiparallel, and becoming exact both at the collision points $x_{j\ell}=0$ and when all $|x_{j\ell}|\to\infty$.
Hence we expect the unknown exact sTG excited state for finite negative $\lambda$ to be well approximated by the state of 
Eq. (\ref{psiFnu}).

To establish a connection with a hard sphere ground state, consider first the case $N=2$. If both atoms have spin up or both have spin down,
then the wave functions are necessarily spatially antisymmetric, vanishing when $x_1=x_2$, so the delta interaction in Eq. (\ref{H})
is ineffective and the energy eigenstates are those of the trapped ideal Fermi gas. On the other hand, if they are antiparallel
in a spin-antisymmetric singlet state arising from an s-wave Feshbach resonance, 
then the ground state wave function is spatially symmetric and of the 
previously-discussed parabolic cylinder function form  $D_\nu(|q|)$ with $q=(x_1-x_2)/x_{\text{osc}}$ \cite{Fra03}. The argument 
establishing the connection with the hard sphere gas in this case is the same as given previously for spinless bosons \cite{GirAst09}. 
The \emph{only} difference between the sTG and HSB wave functions apart from normalization is that
the sTG wave function allows penetration into the region interior $|x|<a_{h.s.}$. This holds for $-\infty<\lambda<0$, 
and the penetration is small when $|\lambda|\gg 1$
so that $x_{\text{node}}=a_{\text{h.s.}}\approx a_{\text{1D}}$. Furthermore, the sTG and hard sphere solutions have the same energy
since they satisfy the same Schr\"{o}dinger equation in the exterior region, and they have the same energy eigenvalue provided that the
energy index $\nu$ of $D_\nu$ \cite{Fra03,GirAst09} is chosen to be the same as that for the ground state solution $D_\nu$ in the entire region 
$|x|\ge 0$ \emph{including} the attractive delta function at the origin. The sTG wave function $D_\nu$
is highly excited relative to the collapsed McGuire state due to the effects of the strong attraction at $x=0$, but when restricted to
the region $|x|\ge x_{\text{node}}$, it is \emph{identical} with the \emph{ground} state of the hard sphere Bose gas apart from
normalization, and the energies of the \emph{highly excited} sTG state and the hard sphere \emph{ground} state are exactly equal 
\cite{Note0}.

A generalization of this theorem 
should hold for all $N>2$. Consider a generalized hard sphere system with a hard core interaction
of diameter $x_{\text{node}}$ for opposite spin pairs, but no interaction for $(\uparrow\uparrow)$ and $(\downarrow\downarrow)$
pairs. For opposite spin pairs, the ground state wave function of this system will vanish when $|x_j-x_\ell|\le x_{\text{node}}$ 
but be nodeless for $|x_j-x_\ell|>x_{\text{node}}$, whereas for $(\uparrow\uparrow)$ and $(\downarrow\downarrow)$ pairs it will
be of the trapped ideal gas form. It therefore seems likely that the unknown exact sTG energy eigenstate, to which  
(\ref{psiFnu}) is an approximation, will differ from such a generalized
hard sphere ground state only in the interior region $|x_j-x_\ell|<x_{\text{node}}$ for opposite spin pairs, and in overall
normalization. Neither the sTG state nor the generalized hard sphere state are known \emph{exactly}. 
Therefore I state the belief of the identity of the sTG and generalized hard sphere wave functions and energies for $N>2$ as a conjecture.

There also exists a quite different sTG state when $N_\uparrow=N_\downarrow=N/2$. In this case it was 
shown recently \cite{CheGuaYinGuaBat09}, 
for a system on a ring with no circumferential trapping potential, that for $-\infty<\lambda\ll -1$ there exists a
gas-like energy eigenstate 
in which all the fermions are tightly bound into $(\uparrow\downarrow)$ dimers which behave as
strongly attracting bosons, forming an exact analog of the untrapped solution for spinless bosons \cite{Chenetal}. 
It is a gas-like sTG state lying much lower than the fermionic sTG state of Eq. (\ref{psiFnu}) due to the negative binding energy of the dimers, but much higher than a McGuire-ideal Fermi hybrid state similar to that of Eq. (\ref{McGuire-ideal Fermi}).
One expects that a similar sTG state
exists for the present case of harmonic trapping. In fact, for $N=2$ it has already been given as the $N_\uparrow=N_\downarrow=1$
case of the previously-discussed exact $N=2$ solution, and for larger even $N$ there will be a Bose-like sTG state 
similar to that of \cite{Chenetal}. However, it could \emph{not} be created by a sudden switch from a strongly repulsive to strongly attractive interaction as in 
\cite{Haletal09}. In fact, the sudden approximation probability of finding the system in such a state after a switch 
$\lambda\gg 1\to -\lambda\ll -1$ is \emph{exactly zero} in the limit $|\lambda|\to\infty$, since the state (\ref{psiFnu})
is an exact energy eigenstate for both $\lambda=+\infty$ and $\lambda=-\infty$. Then by continuity this probability will be $\ll 1$
after a sudden switch $\lambda\gg 1\to -\lambda\ll -1$.
\begin{acknowledgments}
This work is an outgrowth of a talk by and discussion with Hanns-Cristoph N\"{a}gerl at the BEC2009 conference in
Sant Feliu de Guixols, Spain in September 2009.  
I also thank Vladimir Yurovsky for a stimulating conversation there, and Gregory Astrakharchik for comments on this paper. 
It was supported by the U.S. Army Research Laboratory and the U.S. Army Research Office under grant number W911NF-09-1-0228. 
\end{acknowledgments}
\end{document}